# A Criterion to identify maximally entangled nine-qubit state

Xinwei Zha[1,*], Irfan Ahmed[3,4], Da Zhang[2], and Yanpeng Zhang[2,†]

[1]School of Science, Xi'an University of Posts and Telecommunications, Xi'an, 710121, China

[2] Key Laboratory for Physical Electronics and Devices of the Ministry of Education & Shaanxi Key Lab of Information Photonic Technique, Xi'an Jiaotong University, Xi'an 710049, China

[3]Department of Physics, City University of Hong Kong, Kowloon, Hong Kong SAR, China

[4]Electrical Engineering Department, Sukkur IBA University, 65200, Sindh, Pakistan

*zhxw@xupt.edu.cn & †ypzhang@mail.xjtu.edu.cn

**ABSTRACT:** We present a generalized criterion for maximally entangled nine-qubit states, whose minimum averaged subsystem purity should be equal to 1/14. In this note, we prove that absolutely maximally entangled state for nine qubits does not exist. Further, we construct a new genuine nine-qubit maximally entangled states of 110,15 and 1 balanced purities equal to 1/16, , 1/8 and and ¼, respectively. We found that, the marginal density matrices for subsystems of 1,2,3- qubits all completely mixed in these states, therefore, maximally entangled nine-qubit state is a 3-uniform state.

## 1. Introduction

Quantum entanglement plays the most significant role in quantum information processing. Since, the entangled states are quantum information carriers for various computational and information processing tasks suggested by quantum formalism [1,2], the basic role of entanglement in quantum communication within a distant laboratory paradigm is still key discussion [2,3]. The investigation of the entanglement-related phenomena may provide a better understanding of the framework of quantum mechanics [2]. For both theoretical studies and applications, it is imperative to explore, characterize, detect, distillate and quantify all aspects of the quantum entanglement of multipartite quantum systems. Since last decade, a lot of efforts have been made to quantify the amount of entanglement of various multipartite states using sequential coupling of an ancillary system to initially uncorrelated qubits [4]. In particular, investigations have been started from how entangled can two couple get [5] to entanglement monotones [6] and further on to maximally multipartite entangled states [7]. The attention has been extended to multiqubit systems of highly entangled states along with their detection, distribution, structure and quantification [6–11]. Among these bipartite entanglement is well understood but such characterization or classification in multiqubit states is still very challenging.

An $n$-qubit pure state $|\psi\rangle$ is a $k$-uniform state provided that all of its reductions to $k$-qubits are maximally mixed [12,13]. Particularly interesting are those $n$-qubit states which are $[n/2]$-uniform naturally generalized from N-qudit Greenberger-Horne-Zeilinger (GHZ) states. Such states are also called absolutely maximally entangled (AME) states which are a type of pure quantum error correcting codes (QECCs). QECCs are expressed as $((N,K,D))_d$, where $N$ is the length of the code, $K$ is the dimension of the encoding state, $D$ is the Hamming minimum distance, and $d$ is the levels number of the qudit system. Thorough survey of research progress in QECCs has been well classified in the book [14]. Since, AME states have maximal distance separable allowed by the Singleton bound [15], so any violation of this bound in particular is equivalent to a violation of quantum no-cloning theorem.

It is well known that AME states exist only for special values of *n* (n=2,3,5,6). Recently, Felix Huber, et al. [16,17] has proved that there is no AME state for seven qubits. In this note we prove that AME states for nine qubits do not exist and criterion for maximally entangled state of nine-qubit are given. Further, we also construct a maximally entangled nine-qubit states whose marginal density matrices for subsystems of 1, 2 and 3 are completely mixed whereas in case of 4-qubits, most of marginal density matrices, are completely mixed, but not all.

## 2. Criterion of Maximally Entangled nine-Qubit States

Maximally multi-qubit entangled states (MMES) [7] are those states whose entanglement is maximal for every (balanced) bipartition. One can determine whether a state is a MMES when averaged purity of overall [n/2]-qubit subsystems reaches its minimum. The averaged subsystem purity for a state is defined as

$$\pi_{ME} = \binom{n}{n_A}^{-1} \sum \pi_A, \qquad (1)$$

where $n_A = [n/2]$; $\pi_A = Tr_A \rho_A^2$, and $\rho_A = Tr_{\bar{A}} |\psi\rangle\langle\psi|$. Here $Tr_X$ denotes partial trace over subsystem *X*. It is obvious for qubit cases i-e $\frac{1}{N_A} \leq \pi_{ME} \leq 1$, where $N_A = 2^{n_A}$. Generally, a maximally multi-qubit entangled state is a minimizer of $\pi_{ME}$.

The quantity $\pi_{ME}$ measures the average bipartite entanglement over all possible balanced bipartition. From Eq. (1), for nine-qubit states, we have

$$\pi_{ME} = \frac{1}{126}\left(\sum_{i\neq j\neq k\neq l} \pi_{ijkl}\right), \qquad (2)$$

where

$\pi_{ijkl} = Tr_{ijkl}\rho_{ijkl}^2$, $\rho_{ijkl} = Tr_{1\cdots i-1,i+1\cdots j-1, j+1\cdots k-1,k+1\cdots l-1,l+1\cdots n}|\psi\rangle\langle\psi|$, and so on.

In Ref. [17] it has been shown that

$$\pi_{ijkl} = \frac{1}{16} + \frac{1}{16}\left(\sum_u F_u + \sum_{u\neq v} F_{uv} + \sum_{u\neq v\neq w} F_{uvw} + F_{ijkl}\right), \qquad (3)$$
$$(u,v,w) \in (i,j,k,l)$$

where

$$F_i = \sum_{\alpha_1}\langle\psi|\hat{\sigma}_{i\alpha_1}|\psi\rangle^2,$$
$$F_{ij} = \sum_{\alpha_1}\sum_{\alpha_2}\langle\psi|\hat{\sigma}_{i\alpha_1}\hat{\sigma}_{j\alpha_2}|\psi\rangle^2,$$
$$F_{ijk} = \sum_{\alpha_1}\sum_{\alpha_2}\sum_{\alpha_3}\langle\psi|\hat{\sigma}_{i\alpha_1}\hat{\sigma}_{j\alpha_2}\hat{\sigma}_{k\alpha_3}|\psi\rangle^2,$$
$$F_{ijkl} = \sum_{\alpha_1}\sum_{\alpha_2}\sum_{\alpha_3}\sum_{\alpha_4}\langle\psi|\hat{\sigma}_{i\alpha_1}\hat{\sigma}_{j\alpha_2}\hat{\sigma}_{k\alpha_3}\hat{\sigma}_{l\alpha_4}|\psi\rangle^2, \qquad (4)$$
$$\alpha_1,\alpha_2,\alpha_3,\alpha_4 \in (x,y,z).$$

where $\hat{\sigma}_{i\alpha_m}$ $(m=1,2,3,4)$ is the Pauli matrix.

It is obvious that such invariants satisfy $F_i \geq 0$, $F_{ij} \geq 0$, $F_{ijk} \geq 0$, $F_{ijkl} \geq 0$.

Using Eq. (3), for nine-qubit states, we can obtain

$$\pi_{ME} = \frac{1}{16} + \frac{1}{126\times 16}(56C_1 + 21C_2 + 6C_3 + C_4), \qquad (5)$$

where $C_1 = \sum_i F_i$, $C_2 = \sum_{i\neq j} F_{ij}$, $C_3 = \sum_{i\neq j\neq k} F_{ijk}$, $C_4 = \sum_{i\neq j\neq k\neq l} F_{ijkl}$.

For nine-qubit states, we have

$$16\sum_{m=1\cdots i-1,i+1\cdots n} tr\rho_m\tilde{\rho}_m = -18 - 8C_1 - C_2 + C_4, \qquad (6)$$

Using Eq. (6), we can obtain

$$\pi_{ME} = \frac{1}{14} + \frac{1}{16 \times 126}[64C_1 + 22C_2 + 6C_3 + 16 \sum_{m=1\cdots i-1,i+1\cdots n} tr\rho_m\tilde{\rho}_m] \quad (7)$$

Since $C_1 \geq 0$, $C_2 \geq 0$, $C_3 \geq 0$, $tr\rho_m\tilde{\rho}_m \geq 0$ as the reference [18], equation (7) gives a general approach to construct a minimizable structure of $\pi_{ME}$, by which a state with $C_1 = 0$, $C_2 = 0$, $C_3 = 0$, $tr\rho_m\tilde{\rho}_m = 0$ can be minimized to $\pi_{ME} = \frac{1}{14}$. This state with minimization will be realized as a maximally entangled state of nine-qubit. Therefore, a nine-qubit state whose marginal density matrices of 4-qubits are not completely mixed, one can say that absolutely maximally entangled state of nine qubits does not exist.

We can measure the entanglements of the product and GHZ states by this criterion. For the product state, $\sum_i F_i = 1$, $\sum_{i \neq j} F_{ij} = 1$, $\sum_{i \neq j \neq k} F_{ijk} = 1$, $tr\rho_m\tilde{\rho}_m = 0$, $\pi_{ME} = \frac{1}{14} + \frac{13}{14} = 1$. For GHZ states, one can find:

$$\sum_i F_i = 0, \sum_{i \neq j} F_{ij} = 1, \sum_{i \neq j \neq k} F_{ijk} = 0, tr\rho_m\tilde{\rho}_m = \frac{9}{2}, \pi_{ME} = \frac{1}{14} + \frac{6}{14} = \frac{1}{2}.$$

For product stat $\pi_{ME} = 1$ and for GHZ state $\pi_{ME} = \frac{1}{2}$, which suggests that Eq.7 is correct and emphasizing the absolutely maximally entangled of nine qubits state does not exist.

**3. A Maximally Entangled nine-Qubit States**

Construction of genuinely multipartite entangled states is crucial and everyone's problem in the theory of quantum information. Since quantum teleportation, quantum key distribution, dense coding, and QECCs are associated with these genuinely multipartite entangled states. In this regard, here, by performing tedious calculation, we can construct a maximally entangled state of nine-qubit, which can be expressed as

$$|\psi\rangle_{123456789} = \frac{1}{8\sqrt{2}}[(|0000\rangle+|0101\rangle-|1010\rangle-|1111\rangle)_{1234} \otimes \begin{pmatrix} |00000\rangle+|00101\rangle+|01001\rangle+|01100\rangle- \\ |10010\rangle-|10111\rangle+|11011\rangle+|11110\rangle \end{pmatrix}_{56789}$$

$$+(|0001\rangle+|0100\rangle+|1011\rangle+|1110\rangle)_{1234} \otimes \begin{pmatrix} |00011\rangle+|00110\rangle-|01010\rangle-|01111\rangle+ \\ |10001\rangle+|10100\rangle+|11000\rangle+|11101\rangle \end{pmatrix}_{56789} \quad (8)$$

$$+(|0010\rangle-|0111\rangle-|1000\rangle+|1101\rangle)_{1234} \otimes \begin{pmatrix} -|00011\rangle+|00110\rangle-|01010\rangle+|01111\rangle+ \\ |10001\rangle-|10100\rangle+|11000\rangle-|11101\rangle \end{pmatrix}_{56789}$$

$$+(|0011\rangle-|0110\rangle-|1001\rangle+|1100\rangle)_{1234} \otimes \begin{pmatrix} -|00000\rangle+|00101\rangle-|01001\rangle+|01100\rangle- \\ |10010\rangle+|10111\rangle+|11011\rangle-|11110\rangle \end{pmatrix}_{56789} ]$$

By careful calculation, we can obtain, $\pi_{1234} = \frac{1}{4}$;

$$\pi_{1278} = \pi_{1358} = \pi_{1367} = \pi_{1379} = \pi_{1457}$$
$$= \pi_{1679} = \pi_{2368} = \pi_{2458} = \pi_{2479} = \pi_{2489} \quad ;$$
$$= \pi_{2689} = \pi_{3456} = \pi_{3478} = \pi_{3679} = \pi_{4689} = \frac{1}{8}$$

Others

$$\pi_{ijkl} = Tr_{ijkl}\rho_{ijkl}^2 = \frac{1}{16}, ijkl = 1235, 1236, \cdots, 6789 \quad (9)$$

Then, by combining Eq. (2) and (9), $\pi_{ME}$ can be obtained as,

$$\pi_{ME} = \frac{1}{126} \sum_{i \neq j \neq k \neq l} \pi_{ijkl} =$$
$$\frac{1}{126} \times \frac{1}{16} \times 110 + \frac{1}{8} \times 15 + \frac{1}{4} = \frac{1}{14} \quad (10)$$

Thus the nine-qubit states are maximally entangled. Similarly, it is also easy to obtain

$$\pi_{ijk} = Tr_{ijk}\rho_{ijk}^2 = \frac{1}{8}, \quad \pi_{ij} = Tr_{ij}\rho_{ij}^2 = \frac{1}{4}, \quad \pi_i = Tr_i\rho_i^2 = \frac{1}{2} \quad (11)$$

Therefore, for the nine-qubit state, given in Eq. (8), the marginal density matrices for subsystems of 1,2 and 3 are all completely mixed. Therefore, this state is a 3-uniform state.

**3. Discussions and Conclusions**

In conclusion, we have presented a criterion for maximally entangled nine-qubit states, whose minimum averaged subsystem purity should be equal to 1/14and prove that absolutely maximally entangled state of nine qubits does not exist. Based on which a new maximally entangled nine-qubit states was obtained. Such nine-qubit state shows all completely mixed marginal density matrices for 1, 2, and 3-qubits subsystems, which however doesn't apply for all the 4-quibit subsystems. This indicates that such maximally entangled nine-qubit state is a 3-uniform state.

**References**

[1] M. A. Nielsen and I. L. Chuang, *Quantum Computation and Quantum Information* (Cambridge University Press, 2000).

[2] A. Einstein, B. Podolsky, and N. Rosen, Physical Review **47**, 777 (1935).

[3] R. Horodecki, P. Horodecki, M. Horodecki, and K. Horodecki, Reviews of Modern Physics **81**, 865 (2009).

[4] C. Schön, E. Solano, F. Verstraete, J. I. Cirac, and M. M. Wolf, Physical Review Letters **95**, 110503 (2005).

[5] A. Higuchi and A. Sudbery, Physics Letters A **273**, 213 (2000).

[6] A. OSTERLOH and J. SIEWERT, International Journal of Quantum Information **04**, 531 (2006).

[7] P. Facchi, G. Florio, G. Parisi, and S. Pascazio, Physical Review A **77**, 060304 (2008).

[8] J. E. Tapiador, J. C. Hernandez-Castro, J. A. Clark, and S. Stepney, Journal of Physics A: Mathematical and Theoretical **42**, 415301 (2009).

[9] P. Krammer, H. Kampermann, D. Bruß, R. A. Bertlmann, L. C. Kwek, and C. Macchiavello, Physical Review Letters **103**, 100502 (2009).

[10] A. Borras, A. R. Plastino, J. Batle, C. Zander, M. Casas, and A. Plastino, Journal of Physics A: Mathematical and Theoretical **40**, 13407 (2007).

[11] Xin-wei Zha and Chenzhi Yuan and Yanpeng Zhang, Laser Phys. Lett **10**, 045201 (2013).

[12] L. Arnaud and N. J. Cerf, Physical Review A **87**, 012319 (2013).

[13] D. Goyeneche and K. Życzkowski, Physical Review A **90**, 022316 (2014).

[14] D. A. Lidar and T. A. B. Frontmatter, *QUANTUM ERROR CORRECTION-Quantum Error Correction Edited* (Cambridge University Press, Cambridge U.K, 2013).

[15] E. M. Rains, IEEE Transactions on Information Theory **45**, 1827 (1999).

[16] F. Huber, O. Gühne, and J. Siewert, Physical Review Letters **118**, 200502 (2017).

[17] X. wei Zha, I. Ahmed, and Y. Zhang, Results in Physics **6**, 26 (2016).

[18] C. Eltschka and J. Siewert, (2017).